\documentclass[11pt]{iopart}
\usepackage{iopams}
\usepackage{graphicx}
\usepackage[colorlinks,linkcolor=blue,urlcolor=blue,citecolor=blue]{hyperref}
\usepackage{graphicx}
\usepackage{float}
\usepackage{enumerate}
\newcommand{\D}{{\rm d}}

\def\be{\begin{equation}}
\def\ee{\end{equation}}
\def\bea{\begin{eqnarray}}
\def\eea{\end{eqnarray}}

\begin{document}
\title{Effects of refractory period on stochastic resetting}

\date{\today}


\author{Martin R. Evans$^{(1)}$ and Satya N. Majumdar$^{(2)}$}

\address{$^{(1)}$ SUPA, School of Physics and Astronomy, University of
Edinburgh, 
Peter Guthrie Tait Road, Edinburgh EH9 3FD, UK\\
$^{(2)}$ Univ. Paris-Sud, CNRS, LPTMS, UMR 8626, Orsay F-01405, France}

\ead{m.evans@ed.ac.uk, satya.majumdar@u-psud.fr}

\begin{abstract}

We consider a stochastic process undergoing resetting after which a random refractory period is imposed. In 
this period the process is quiescent and remains at the resetting position. Using a first-renewal approach, 
we compute exactly the stationary position distribution and analyse the emergence of a delta peak at the 
resetting position. In the case of a power-law distribution for the refractory period we find slow relaxation. We generalise our results to the case when the resetting period and the refractory period 
are correlated, by computing the Laplace transform of the survival probability of the process and the mean 
first passage time, i.e., the mean time to completion of a task. We also compute exactly the joint distribution of the 
active and absorption time to a fixed target.

\end{abstract}

 \pacs{05.40.−a, 02.50.−r, 87.23.Ge, 05.10.Gg}

\vspace{2pc}
\noindent{\bf Keywords}: stochastic resetting, refractory period, diffusion\\[1ex]

\noindent Accepted for {\it Journal of Physics A: Mathematical and Theoretical Letters}


\section{Introduction} 
Resetting a stochastic process may improve the time to complete a task
considerably.  For example if the task at hand is to locate a fixed target,
a purely diffusive process has infinite mean first passage time (MFPT)
to the target. However, when a resetting process is introduced the MFPT
is rendered finite \cite{EM11a}.  The idea of resetting a random
process to improve efficiency  or throughput
extends to diffusion processes with random target position~\cite{EM11b} and
to arbitrary dimensions~\cite{EM14}, non-diffusive processes such as L\'evy flights~\cite{KMSS14}
and active run and tumble particles~\cite{EM18},
to computer algorithms \cite{VAVA91,LSZ93,TFP08},
Michaelis-Menten type reaction schemes \cite{RUK14,RRU15,R16} and kinetic 
proofreading mechanisms \cite{TBZL02,MHL12,MHL14}. The central idea is 
that resetting may eliminate a long tail of completion times 
in the absence of resetting.

However, in most contexts resetting does not come without a cost.  For example, 
`teleporting' a searcher back to where he started must take some time and effort 
\cite{AG13,BMSV07}. In a different context, neuron and nerve cells undergo 
resetting events involving the firing of action potentials after which the cell is in 
a quiescent state in which subsequent stimuli are ineffective. Such a {\em 
refractory} period may be absolute, in which case a subsequent action potential 
cannot occur, or relative in which case a 
subsequent action potential is inhibited but not impossible. Refractory periods may 
be psychological as well as physiological in nature. For example, large movements in 
the stock exchange are often interpreted as resetting events after which there is 
often a slow-down period in which activity abates \cite{sornette}.

A recent series of papers \cite{RUK14,RRU15,R16} has considered a Generalised Michaelis-Menten chemical 
reaction scheme in which there are three stages: i) unbound enzyme and substrate; ii) enzyme bound to 
substrate; iii) the catalytic production of a product. The cycle goes from i) to ii) when the enzyme binds 
with a certain rate $k_{\rm on}$ and reverts to i) when the enzyme unbinds with rate $k_{\rm off}$; the cycle 
goes from ii) to iii) through a complicated possible series of reactions that result in a waiting time 
distribution $f(t)$ for the production of the product.
 
Remarkably, as noted in \cite{RUK14,RRU15,R16}, this reaction scheme maps precisely onto a  
general resetting process
(the unbinding which occurs with rate $k_{\rm off}$) which then incurs a refractory period 
corresponding to the binding time controlled by rate $k_{\rm on}$.
The waiting distribution $f(t)$ characterises the  stochastic process which is reset.
The works \cite{RUK14,RRU15} (see also \cite{KCMEX05}) have derived important results for the mean time 
for the production of a product and the  
Laplace transform of the distribution of the production time. Moreover, it has been shown that 
interesting universality emerges when the resetting  process is chosen to optimise the time to 
completion of the process \cite{R16}.

In this note we study and extend further the effects of the presence of a non-zero refractory period on
other properties of a generic process undergoing stochastic resetting, going beyond the first-passage time
distribution. 
More precisely,  
We consider 
a stochastic process $x(t)$ in
continuous time $t$
under resetting, wherein immediately after a
resetting event we impose a refractory period:
during a refractory period, the particle
`sleeps' (no movement or activity, i.e., a quiescent state). The duration of the refractory period,   
$\tau$,  is taken to be a random variable, drawn
independently after each reset event from a distribution
$W(\tau)$. The distribution $W(\tau)$ may or may not have finite mean.
At the end of the random refractory period, the process resumes its
random motion till the next reset event. The evolution is shown
schematically in Fig. 1, where the reset position $X_r$ is taken to be
the origin $x=0$ for simplicity.
We shall initially consider the case where the resetting is Poissonian i.e. 
occurs with fixed rate but later generalise to arbitrary joint distribution for the 
resetting period  and the subsequent refractory period.

We consider three novel effects of a refractory period on resetting.
(i) We first derive in Section 2.1 the stationary state for the case of Poissonian reset
(where resetting occurs with constant rate $r$), but an arbitrary distribution $W(\tau)$ for  
$\tau$, the refractory period after the reset.
We show how the stationary state contains a delta peak at the resetting position corresponding to the  
process in the quiescent state during its refractory period.
In section 2.2 we study the long-time relaxation of this delta peak
(ii) In section 3 We then go on to consider correlations between the refractory period and the period of activity before the  
reset that preceded the refractory period.
For example, it would be natural 
if after a long period of activity before the reset the refractory period were longer than after a short 
period of activity.
We show that the stationary state depends on these correlations in a {\it weak} way, i.e., it depends only
on the marginal distribution of the first resetting time (which  may depend on the correlations).
However the survival probability of the process in the presence of an absorbing target is {\it strongly} affected 
by the correlation
and we compute explicitly the Laplace transform of the survival probability, from which we 
obtain the mean first passage time under the reset. 
(iii) Finally, in Section 4, for a process with Poissonian resetting and independent refractory period, we compute exactly the 
joint distribution of the survival time and the active time till the target is found,
i.e.  the time during which the process is not in the quiescent state.
This quantity is of interest since in an optimisation problem it may be appropriate  
to consider the active time as distinct from time spent in the quiescent state.

\section{Renewal equation for the probability distribution}
For simplicity we begin with the established  
case of Poissonian resetting and refractory period probability density $W(\tau)$
(see \cite{RUK14,RRU15,R16} although, as noted above, the notation is slightly different).  Between resets 
we follow a general stochastic process $x(t)$.
We denote by $G_0(x,t)$  the time-dependent probability density   for this process to have reached $x$ after time $t$ (in the absence of resetting) where we have suppressed the initial condition which we take to be coincident with  the resetting position $X_r=0$ (taken here to be at the origin).
We shall refer to $G_0(x,t)$ as the propagator of the process. 

We begin by writing a renewal equation for the probability density which
uses the {\rm first reset} to renew the process. 
Related first-renewal equations have been previously used in the case of  
the Generalised Michaelis-Menten reaction scheme \cite{RUK14,RRU15,R16}
and non-Poissonian reset \cite{PKE16}.
In other contexts last-renewal equations have been employed \cite{EM11a,MSS15,KMSS14,EM18}.  
Recently, a unified renewal approach has been developed \cite{PR17,HK16,CS18}.

First note that for a Poissonian  resetting process (constant rate $r$) the probability of no resets occurring up until time $t$ is given by $ {\rm e}^{-rt}$ and the probability density for the first reset to occur at time $t_1$ is
$r {\rm e}^{-rt_1}$. Then we may  write down
an equation for the probability density for the process to be at $x$ at time $t$.
\begin{eqnarray}
P(x,t) &=& {\rm e}^{-rt} G_0(x,t)\nonumber \\
&+& 
r \int_0^t \D t_1 {\rm e}^{-rt_1}\int_0^{t-t_1} \D \tau_2 W(\tau_2) P(x,t-t_1-\tau_2)
\nonumber \\
&+& r \int_0^t \D t_1 {\rm e}^{-rt_1} \int_{t-t_1}^\infty \D \tau_2 W(\tau_2) \delta(x)\;.  \label{renew1}
\end{eqnarray}
We note here that, since our refractory period occurs after a reset, 
the initial condition is different from that considered in the series of papers
\cite{RUK14,RRU15,R16} where the process starts off with a refractory period.

\begin{figure}[t]  
\centering
\includegraphics[scale=0.4]{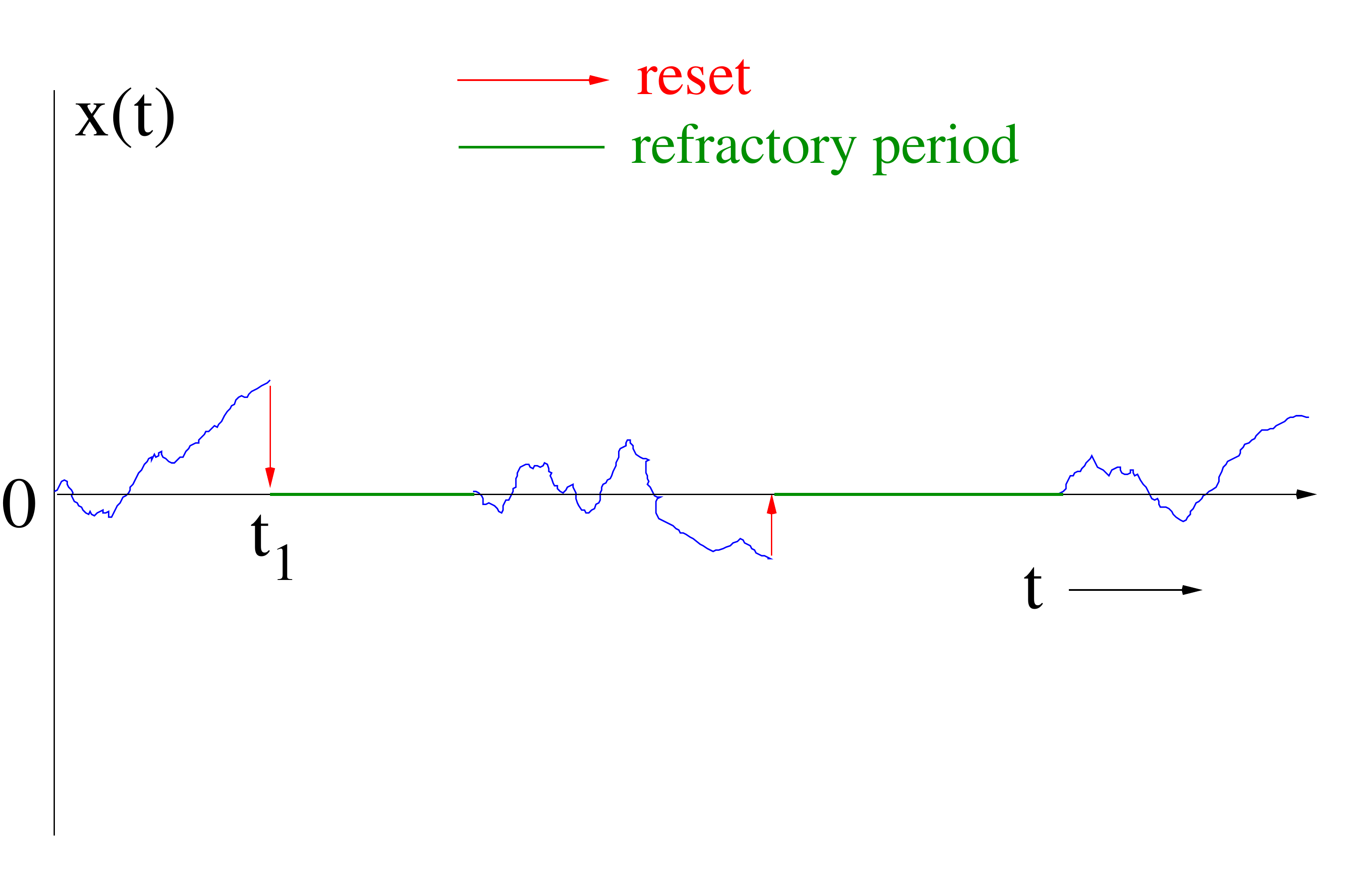}
\caption{Schematic evolution of a stochastic process $x(t)$ as a function of time $t$, 
starting from $x(0)=0$.
The red arrows denote the reset (to the origin) events. 
The first reset happens at time $t_1$.
Immediately following a reset,
the particle spends a random refractory period during which
the particle does not move (shown by the 
think green lines). The process resumes its dynamics only at the end of the refractory
period.
}
\label{fig1_refrac}
\end{figure}

The first term on the right hand side (rhs) of (\ref{renew1}) is the contribution from trajectories in which there is no resetting, which occurs with probability ${\rm e}^{-rt}$; the second term integrates over the first resetting time $t_1$,
and the subsequent refractory period with duration $\tau_2$
which finishes before $t$ leaving a period $t-t_1- \tau_2$
for the particle to pass from the resetting site to $x$;
the third contribution is
from  first resets at $t_1$ with refractory periods $\tau_2$ which last until beyond time $t$, leaving the particle at the resetting position (here the origin).

We now take the Laplace transform of (\ref{renew1})
which is defined as
\begin{equation}
\tilde P(x, s) = \int_0^\infty \D t\, {\rm e}^{-st} P (x,t)\;.
\end{equation}
The Laplace transform of the first term yields
$\tilde G_0(x,r+s)$ which is the Laplace transform (with Laplace variable $r+s$) of the propagator.
The Laplace transform of the  second  term yields from the convolution theorem
$\displaystyle \frac{r}{r+s} \tilde W(s) \tilde P(x,s)\;,$
where $\tilde W(s)$ is the Laplace transform of the refractory period distribution, and the Laplace transform of the  third   term yields
\begin{eqnarray}
&&\delta(x)r \int_0^\infty \D t\, {\rm e}^{-st}
\int_0^t \D t_1 {\rm e}^{-rt_1}
 \int_{t-t_1}^\infty \D \tau_2 W(\tau_2)\nonumber \\
&=& \delta(x)r \int_0^\infty \D t_1 {\rm e}^{-rt_1}
\int_0^\infty \D \tau_2  W(\tau_2) \int_{t_1}^{\tau_2 + t_1} \D t  {\rm e}^{-st}
\nonumber \\
&=&
\delta(x)\frac{r}{s(r+s)}\left[1- \tilde W(s)\right]\;.
\label{lt2}
\end{eqnarray}
Putting these results together yields
\begin{equation}
\tilde P(x,s) = \frac{1}{r+s - r \tilde W(s)}
\left[(r+s) \tilde G_0(x,r+s)
+ \delta(x)\frac{r}{s}\left[1- \tilde W(s)\right]\right]\;.
\label{Plt}
\end{equation}
As a check, we may integrate Eq. (\ref{Plt}) over $x$, using $\int_{-\infty}^{\infty} \tilde G_0(x,s) dx=1/s$.
We then recover $\int_{-\infty}^{\infty} \tilde P(x,s) dx=1/s$ as expected, since $P(x,t)$ is normalised to unity.
Another useful check concerns the
the case of vanishing refractory period
$W(\tau) = \delta(\tau)$. Then  $\tilde W(s) =1$ and we find
\begin{equation}
\tilde P_{\tau =0}(x,s) \to  \frac{(r+s)}{s}
\tilde G_0(x,r+s)\;.
\label{Plttau0}
\end{equation}
For instance, for a simple diffusing particle on a line with diffusion constant $D$ and starting at the origin 
$x=0$, the bare propagator is
$G_0(x,t)= e^{-x^2/{4Dt}}/\sqrt{4\pi Dt}$. The Laplace transform with respect to $t$ gives,
${\tilde G_0}(x,s)= e^{- \sqrt{s/D}\, |x|}/\sqrt{4\,D\,s}$. Hence, in this case, Eq. (\ref{Plttau0})
gives
\begin{equation}
\tilde P_{\tau =0}(x,s)= \frac{1}{s}\, \sqrt{\frac{r+s}{4D}}\, e^{- \sqrt{\frac{r+s}{D}}\, |x|}\, 
\label{Pltttau0_diff}
\end{equation}
which is completely consistent with the full time-dependent solution in the absence of a refractory period~\cite{EM14,MSS15}.

\subsection{Stationary State}

The stationary state $P^{\rm st}(x)$ is easy to extract as the coefficient of $1/s$
in the small $s$ expansion of (\ref{Plt}).
In the case of vanishing refractory period (\ref{Plttau0})
we recover $P^{\rm st}(x) = r \tilde G_0(x,r)$ \cite{EM11a}.
In  the general case, assuming $W(\tau)$ has finite mean $\langle \tau \rangle$, we have for small $s$
\begin{equation}
\tilde W(s) \simeq 1 - s \langle \tau \rangle\;.
\end{equation}
Then we obtain the stationary distribution
\begin{equation}
\label{Pstat.1}
P^{\rm st}(x) = \frac{r}{1+ r \langle \tau \rangle}
\left[\tilde G_0(x,r)
+ \delta(x)\,\langle \tau \rangle \right]\;.
\end{equation}
Thus the probability distribution is a superposition of 
the distribution with zero refractory period (which is given by
$\displaystyle r\tilde G_0(x,r)$)
and
a delta function at the resetting position with relative weight equal to 
$r\langle \tau \rangle$ i.e. the ratio of the mean refractory period to the mean period between resets.
We shall see in Section 3 that the form of (\ref{Pstat.1}) is robust to the introduction of correlations between refractory period and resetting period.

On the other hand in the case where the mean of $W(\tau)$ diverges
we find that the stationary state  
$P^{\rm st}(x) \to \delta(x)$, i.e., the stationary state  is dominated by 
the divergent mean refractory period.
  
\subsection{Long-time relaxation}
We now consider the long-time relaxation to the stationary state. 
In the absence of a refractory period, the relaxation to the stationary
state of a stochastic process in presence of resetting was analysed in Ref. \cite{MSS15}
for diffusion on a line, and an interesting dynamical phase transition was found.
Here, we will focus on the relaxation of the new feature in $P(x,t)$ that is induced
entirely by the presence of a refractory period. More specifically, we
note from Eq. (\ref{Plt}) that
the delta peak at the resetting position $x=0$, induced by the refractory period, 
is present at all times, and not just in the stationary state. 
In this section, we focus on the growth of this delta peak in (\ref{Plt}). If one
inverts the Laplace transform in Eq. \ref{Plt}, the delta peak $\delta(x)$ will
have a prefactor which we denote $B(t)$.  From Eq. (\ref{Plt}), the Laplace transform of $B(t)$ reads
\begin{equation}
\tilde B(s) = 
\frac{r\left[1- \tilde W(s)\right]}{s(r+s - r \tilde W(s))}\;.
\label{BS_def}
\end{equation}
The late time behaviour is determined by the singularities
of $\tilde B(s)$ with largest real part. 

To illustrate the case in which $W(\tau)$ has finite mean, let  us consider
an exponential distribution $W(\tau) = a {\rm e}^{-a\tau}$
where $a = 1/\langle \tau \rangle$. Then after easy computation we obtain
\begin{equation}
\tilde B(s) = 
\frac{r}{s(s + a+ r)}
\end{equation}
and inversion of this expression yields
\begin{equation}
B(t) = 
\frac{r}{(a+ r)}\left( 1- {\rm e}^{-(a+r)t}\right)
\;.
\end{equation}
Thus the relaxation time $\tau_{rel} = (a+r)^{-1}$ for the delta peak is given by half the harmonic mean of the 
mean resetting time  and the exponential decay scale
of the  refractory period. We expect a similar relaxation 
scenario to hold whenever $W(\tau)$ has an exponential tail.

We next consider the case when the distribution of the refractory period
$W(\tau)$ has a power law decay for large $\tau$, but still with a finite first moment 
\begin{equation}
W(\tau)\simeq \frac{A}{\tau^{\gamma}}\,\,\, \gamma>2\;,
\label{Wtau.1}
\end{equation}
where $A$ is a constant and 
the exponent $\gamma>2$ ensures that $\langle \tau\rangle$ is finite. In this case, the small
$s$ behaviour of the Laplace transform is given by (for a related analysis in the context of mass transport
models, see Ref.~\cite{EMZ06})
\begin{equation}
\tilde W (s)= \sum_{k=0}^{n-1} (-1)^k \frac{\mu_k}{k!}\, s^k + b\, s^{\gamma-1}+\ldots 
\label{WS_zrp}
\end{equation}
where $\mu_0=1$, $\mu_1=\langle \tau\rangle$, $b= A\, \Gamma(1-\gamma)$ and $n= {\rm int}(\gamma)$. We 
consider, for simplicity, the case for non-integer $\gamma>2$. For integer $\gamma>2$, there are additional 
logarithmic corrections~\cite{EMZ06}. Substituting this small $s$ behaviour of $\tilde W(s)$ in Eq. 
(\ref{BS_def}), we get for small s 
\begin{equation} \tilde B(s)= \frac{r\langle \tau\rangle}{1+r\, \langle 
\tau\rangle}\, \frac{1}{s} - \frac{b r}{(1 +r \langle \tau\rangle)^2}\, s^{\gamma-3} + {\rm analytic}\,\,{\rm 
terms}\,  
\label{BS.1} 
\end{equation}
where we have kept the most dominant nonanalytic term in the small $s$ expansion. 
The first term on the rhs, when Laplace inverted, will 
lead to the stationary value at late times. The next subleading non-analytic term, when inverted, gives 
rise
to a temporal power-law decay to the stationary state. The other analytic terms correspond
to faster than power-law decay at late times. Inverting explicitly the first two terms in Eq. (\ref{BS.1})
then leads to the late time result (for noninteger $\gamma>2$)
\begin{equation}
B(t) \simeq \frac{r\langle \tau\rangle}{1+r\, \langle \tau\rangle} - \frac{1}{(\gamma-2)(\gamma-1)}\, 
\frac{A\, r}{(1+r \langle \tau\rangle)^2}\, \frac{1}{t^{\gamma-2}}\, ,
\label{Bt.gamma>2}
\end{equation}
where $A$ is defined in Eq. (\ref{Wtau.1}), and we have used
${\cal L}_s^{-1}[s^{p}]= t^{-1-p}/{\Gamma(-p)}$.
Thus, the late time relaxation to the stationary value is algebraic with an exponent $\gamma-2$. A
similar analysis can be performed for the case when $\gamma>2$ is an integer.

As an example of a case where the mean refractory period $\langle \tau\rangle$
diverges, we again consider $W(\tau)$ with an algebraic tail as in Eq. \ref{Wtau.1}, but with
exponent $1<\gamma<2$, i.e, 
The Laplace transform of $W(\tau)$ behaves for small $s$ as in Eq. (\ref{WS_zrp}), except that
now $n={\rm int}(\gamma)=1$, hence the first two leading terms predict 
\begin{equation}
\tilde W(s) \simeq 1 + b\, s^{\gamma-1} +\ldots  \quad\quad{\rm with}\,\,\,  1<\gamma<2
\label{WS_meand}
\end{equation}
where the constant $b= A\, \Gamma(1-\gamma)$.
Thus, from Eq. (\ref{BS_def}), we find for small $s$,
\begin{equation}
\tilde B(s) \simeq  \frac{1}{s} + \frac{1}{br}\, s^{1-\gamma}\;,
\end{equation}
which, upon inversion, yields the asymptotic behaviour for $1<\gamma<2$
\begin{equation}
 B(t) \simeq  1  + \frac{(\gamma-1)\sin(\pi \gamma)}{A\,r\,\pi}\, \frac{1}{t^{2-\gamma}}\, .
\label{Bt.gamma<2}
\end{equation}

Comparing Eqs. (\ref{Bt.gamma>2}) for $2<\gamma$ and (\ref{Bt.gamma<2}) for $1<\gamma<2$, 
we see that in both cases
the amplitude $B(t)$ of the delta peak at $x=0$ approaches its stationary value algebraically as 
$t^{-\theta}$ at late times,
where the exponent $\theta= \gamma-2$ for $\gamma>2$ and $\theta=2-\gamma$ for $1<\gamma<2$.
Interestingly, the exponent $\theta$ approaches $0$ as $\gamma\to 2$ from either side, indicating
the slowest relaxation exactly for $\gamma=2$. Indeed, this marginal case $\gamma=2$ can also
be worked out explicitly. In this case, $W(\tau)\simeq A/\tau^2$ for large $\tau$, and 
consequently, for small $s$, 
$\tilde W(s)\simeq  1 + A\, s\, \ln(s)$. Substituting this behaviour in Eq. (\ref{BS_def}), we get
$\tilde B(s)\simeq 1/s + 1/[A\, s\, \ln s]$. Upon inverting, we obtain, for $\gamma=2$
\begin{equation}
B(t) \simeq 1 - \frac{1}{A\, \ln t}
\label{Bt.gamma2}
\end{equation}
thus demonstrating an ultra-slow inverse logarithmic relaxation to the stationary state in this marginal case
$\gamma=2$. Such inverse logarithmic relaxation was found previously in a completely different context of
an Ising-Glaber chain relaxing in the presence of kinetic disorder~\cite{MDG01}.

\section{Correlated resetting and refractory period distribution}
In the previous section the time $t_1$ to the first reset and the 
duration $\tau_2$ of the subsequent refractory 
period were independent random variables (see equation (\ref{renew1})). It is of interest 
to consider the more general situation where they are correlated. For example, it would be natural 
if after a long period of activity before the first reset the refractory period were longer than after a short period of activity.

Generally we can consider the joint probability density $H(t,\tau)$ for the the first reset event to occur at time $t$ and the subsequent refractory period to be of duration $\tau$.
The marginal distributions for the time of the first reset event and the refractory period duration are
\begin{eqnarray}
h(t) &=& \int \D \tau H(t,\tau) \label{hdef}\\ 
W(\tau) &=& \int \D t H(t,\tau)\label{Wdef}\;.
\end{eqnarray}
We also require the probability $g(t)$ of no resetting up to time t which is
given by
\begin{equation}
g(t) = \int_t ^\infty h(t') \D t'\;.
\label{gdef}
\end{equation}

\subsection{Examples of correlated resetting period/refractory period distributions}
\label{sec:ex}
As an example let us consider the case
\begin{equation}
H(t, \tau) = r {\rm e}^{-rt} \frac{1}{a_1t} {\rm e}^{-\tau/(a_1t)}\;.
\end{equation}
Thus the resetting period $t$ and the refractory period $\tau$ are both  exponentially distributed but the mean of $\tau$ for given $t$ increases linearly with the time $t$ to the reset. 
In this case we can compute explicitly the marginal distributions 
\begin{eqnarray}
h(t) &=& r {\rm e}^{-rt} \\ 
\displaystyle W(\tau) &=& \frac{2r}{a_1} K_0(2^{3/2}(\tau r/a_1)^{1/2})
\end{eqnarray}
where $K_0$ is the modified Bessel function of the second kind.
Note that $W(\tau)$ decays for large $\tau$ as 
$\displaystyle W(\tau) \simeq \left( \frac{r^3}{2\pi^2a_1^3\tau}\right)^{1/4}{\rm e}^{-2^{3/2} (\tau r/a_1)^{1/2}}$.
Thus there is a slow stretched exponential decay of the refractory period distribution.

\subsection{Stationary distribution}
Following the first-renewal approach  presented in Section 2  it is 
easy to arrive at the Laplace transform for the probability distribution
with resetting to the origin
\begin{equation}
\tilde P(x,s) = \frac{1}{1-\tilde H(s, s)}
\left[ \int_0^\infty \D t\, {\rm e}^{-st} g(t) G_0(x,t)
+ \delta(x)\frac{1}{s}\left[\tilde H(s,0)- \tilde H(s,s)\right]\right]\;,
\label{Pltgen}
\end{equation}
where we define the double Laplace transform 
\begin{equation}
\tilde H(s, m) = \int_0^\infty \D t \int_0^\infty \D \tau
{\rm e}^{-st- m \tau}H(t, \tau)\;.
\end{equation}
As before the stationary state $P^{\rm st}(x)$ is extracted as the coefficient of $1/s$
in the small $s$ expansion of (\ref{Pltgen})
and 
in the case where the time to first reset and refractory period have finite means $\langle R \rangle$ and $\langle \tau \rangle$, we obtain
the stationary distribution
\begin{equation}
P^{\rm st}(x) = \frac{1}{ \langle R \rangle + \langle \tau \rangle}
\left[ \int_0^\infty \D t\, g(t) G_0(x,t)
+ \delta(x)\langle \tau \rangle \right]\;.
\label{Pstgen}
\end{equation}
 Interestingly, the stationary distribution $P^{\rm st}(x)$ depends 
on the joint distribution of the first resetting time and the refractory period $H(t,\tau)$ only through the marginal
$h(t)$ of the first resetting time
(through $g(t)=\int_t^{\infty}h(t')dt'$ in Eq. (\ref{gdef})). The correlation between $t$ and $\tau$ present
in the joint distribution $H(t,\tau)$ may  determine the marginal $h(t)= \int_0^{\infty} H(t,\tau)\, d\tau$, and
thereby the stationary state. However, the dependence of the stationary distribution on the correlation
is {\it weak} in this sense.
Note that taking $\langle \tau \rangle =0$ we obtain the stationary distribution without
the refractory period 
\begin{equation} P^{\rm st}_{\tau=0}(x) = \frac{1}{\langle R \rangle}
 \int_0^\infty \D t\, g(t) G_0(x,t)\;,
\end{equation}
see \cite{PKE16}.
Again we see from (\ref{Pstgen}) that a finite refractory period introduces a delta function at the resetting position with relative weight equal to the ratio 
of mean refractory period to mean resetting period.

\subsection{Survival Probability}

Let us now consider 
the process
under resetting to $X_r$ followed by a refractory period and with the addition of 
an absorbing boundary at the origin.
The process is absorbed when it reaches the origin for the first time.
In the context of a search process we can think of the time to absorption as the time to 
find a target located at the orgin. More generally we can consider the time to reach the origin as the 
time to completion of some task.
In the following we will refer to the survival probability
and the mean first passage time (MFPT)  of the  process but with the general context of 
random time to completion of a task in mind.

We again  take advantage of a renewal equation.
We first define
$Q_r (x_0,t)$  as the survival probability of the 
process in the {\em presence} of resetting and refractory period and $Q_0 (x_0,t)$ 
as the survival probability in the {\em absence} of resetting, both starting from $x_0$ at $t=0$.
Note that we do not keep track of the final position of the particle and in fact, the
final position has already been integrated over in the definition of the survival probability.
It is well known that for the computation of the first-passage 
probabilities for stochastic processes, direct calculation of the survival probability 
treating the starting position $x_0$
as a variable (the so called backward approach)
is much more advantageous~\cite{bfp_review,pers_review}, 
compared to the ones where one keeps track of the final position and then
integrates over it.
Note that the initial position $x_0$, treated as a variable here, 
may be set equal to the resetting position $X_r$ at the end of the calculation.

We may write down  a renewal equation using similar reasoning to that of  (\ref{renew1})
\begin{eqnarray}
Q_r(x_0,t) &=& g(t) Q_0(x_0,t)\nonumber\\
&+& 
 \int_0^\infty\D t_1   Q_0(x_0, t_1)
\int_{0}^{\infty} \D \tau_2 H(t_1,\tau_2)\int_0^\infty \D t_3  Q_r(X_r,t_3) \delta(t-(t_1+\tau_2+t_3))\nonumber \\
&+&
 \int_0^t\D t_1 
\int_{t-t_1}^{\infty} \D \tau_2  H(t_1, \tau_2) Q_0(x_0, t_1) \;. \label{Qrenew}
\end{eqnarray}
The first term on the rhs is the contribution from survival trajectories
up to time $t$ in which there is no resetting, which occurs with probability $g(t)$. The second term on the rhs of (\ref{Qrenew}) integrates the
contributions from survival trajectories in which the first reset occurs at time $t_1$ and the refractory period $\tau_2$ ends before time $t$, which has probability density $H(t_1, \tau_2)$; these trajectories have survival probability $Q_0(x_0, t_1)$ for the initial period until the first reset and
survival probability $Q_r(X_r, t_3)$ from the end of the refractory period at $t_1+\tau_2$ until final time $t$.
The third term on the rhs of (\ref{Qrenew}) integrates the
contributions from trajectories which survive without resetting until time
$t_1$ at which  the first reset occurs,  and for which the subsequent  refractory period $\tau_2$ is ongoing at time $t$.

We now  take  the  Laplace transform with  Laplace variable $s$  of
(\ref{Qrenew}). 
The Laplace transform of the  third   term yields
\begin{eqnarray}
&& \int_0^\infty \D t\, {\rm e}^{-st}
\int_0^t \D t_1 
 \int_{t-t_1}^\infty \D \tau_2\, H(t_1,\tau_2) Q_0(x_0, t_1)\nonumber \\
&=& \int_0^\infty \D t_1 
\int_0^\infty \D \tau_2 \int_{t_1}^{t_1+\tau_2} \D t\,  {\rm e}^{-st}
H(t_1,\tau_2) Q_0(x_0, t_1) \nonumber \\
&=& \int_0^\infty \D t_1 \frac{1}{s}\left[
{\rm e}^{-st_1} - {\rm e}^{-s(t_1+\tau_2)}\right]
H(t_1,\tau_2) Q_0(x_0, t_1) \nonumber  \\
&=& \frac{1}{s} \left[ \int_0^\infty \D t_1 
{\rm e}^{-st_1}
h(t_1)  Q_0(x_0, t_1)  - 
\int_0^\infty \D t_1  \int_0^\infty \D \tau_2 
{\rm e}^{-s(t_1+\tau_2)}
H(t_1,\tau_2) Q_0(x_0, t_1)\right] \nonumber 
\label{lt3}\;.
\end{eqnarray}
Putting this result together with computations similar to those used to obtain (\ref{Plt}), we find
\begin{eqnarray}
\tilde Q_r(x_0,s) &=&  \int_0^\infty\D t\, {\rm e}^{-st}g(t) Q_0(x_0,t) \\
 &+&  \tilde Q_r(X_r,s) \int_0^\infty\D t_1 \int_0^\infty\D \tau_2\, {\rm e}^{-s(t_1+\tau_2)}
H(t_1,\tau_2) Q_0(x_0, t_1) \nonumber \\
&+& \frac{1}{s} \left[ \int_0^\infty \D t_1 
{\rm e}^{-st_1}
h(t_1)  Q_0(x_0, t_1)  - 
\int_0^\infty \D t_1  \int_0^\infty \D \tau_2 \,
{\rm e}^{-s(t_1+\tau_2)}
H(t_1,\tau_2) Q_0(x_0, t_1)\right] \;. \nonumber 
\end{eqnarray}
Setting $x_0 = X_r$ and rearranging yields
\begin{eqnarray}
\tilde Q_r(X_r,s)
&=& \left[1 - \int_0^\infty \D t \int_0^\infty \D \tau\, {\rm e}^{-s(t+\tau)} H(t,\tau) Q_0(X_r,t)\right]^{-1} \nonumber\\
&\times& \left \{ 
 \int_0^\infty\D t\, {\rm e}^{-st}g(t) Q_0(X_r,t) \right. \\ 
&&\left. + \frac{1}{s} \left[ \int_0^\infty \D t\, {\rm e}^{-st}
h(t)  Q_0(x_0, t)  - 
\int_0^\infty \D t  \int_0^\infty \D \tau \,
{\rm e}^{-s(t+\tau)}
H(t,\tau) Q_0(X_r, t)\right] \right\}\;. \nonumber
\label{Qrs}
\end{eqnarray}
This is the main result of this section.  It gives the Laplace transform of the survival probability under resetting with refractory period in terms 
of the Laplace transforms of the refractory period  distribution  and  survival 
probability without resetting.
It generalises 
a previous result for the case of uncorrelated refractory period and
resetting period \cite{RUK14,KCMEX05, RRU15,R16} (although note that the initial condition in those works is  different to ours leading to slightly different expressions).
To check the case of uncorrelated resetting period and refractory period we can set
$H(t, \tau)= h(t) W(\tau)$  and obtain
\begin{eqnarray}
\tilde Q_r(X_r,s)
&=& \left[1 - \tilde W(s) \int_0^\infty \D t\,{\rm e}^{-st} h(t) Q_0(X_r,t)\right]^{-1} \nonumber\\
&\times& \left \{ 
 \int_0^\infty\D t\, {\rm e}^{-st}g(t) Q_0(X_r,t) \right. \\ 
&&\left. + \frac{1- \tilde W(s)}{s} \int_0^\infty \D t\, {\rm e}^{-st}
h(t)  Q_0(x_0, t)   \right\} \;. \nonumber
\label{Qrsuncorr}
\end{eqnarray}
We can further check the vanishing refractory period case where $\tilde W(s) =1$ in which case we recover 
the result \cite{PKE16} 
\begin{eqnarray}
\tilde Q_r(X_r,s)
&=& \frac{ \int_0^\infty\D t\, {\rm e}^{-st}g(t) Q_0(X_r,t)         }{1 -
\int_0^\infty \D t\, {\rm e}^{-st}
h(t)  Q_0(x_0, t)
}\;.
\label{Qrs.ref0}
\end{eqnarray}

\subsection{Mean first passage time}
The mean first passage time to the origin (or equivalently the mean time to absorption at the origin), $T(X_r)$, is conveniently given by
\begin{equation}
T(X_r) = \tilde Q_r(X_r, s\to 0)\;.
\end{equation}
In the $s\to 0$ limit it can be checked that (\ref{Qrs}) reduces to
\begin{eqnarray}
T(X_r)   &=&  \frac{\int_0^\infty \D t\, g(t)  Q_0(X_r,t)
+ \int_0^\infty \D t 
\langle \tau | t \rangle h(t) Q_0(X_r,t)}
{1- \int_0^\infty \D t\, h(t)  Q_0(X_r,t)}\;.
\label{Tresgen}
 \end{eqnarray}
where $\langle \tau | t \rangle = \int_0^\infty \D \tau\,  \tau H(t,\tau)/h(t)$
is the average of the refractory period conditioned on the resetting period $t$.

In the zero refractory period limit the second term in the numerator vanishes and one 
recovers the result  of \cite{PKE16} 
\begin{eqnarray}
T_{\tau =0}(X_r)   &\to &  \frac{\int_0^\infty \D t\, g(t)  Q_0(X_r,t) }
{1- \int_0^\infty \D t\, h(t)  Q_0(X_r,t)}\;.
\end{eqnarray}
Thus (\ref{Tresgen}) shows that the mean time to absorption is that for the case of zero refractory period  plus a second term  coming from the occurrence of a refractory period. To understand this term better
we note first that the
quantity
$\int_0^\infty \D t\, h(t)  Q_0(X_r,t)$ is the probability  of survival of the process up to the first reset.
Thus, the 
denominator of (\ref{Tresgen})  $1- \int_0^\infty \D t\, h(t)  Q_0(X_r,t)$ is 
the complementary probability of absorption of the process before the first reset occurs.
Now consider the term in the numerator of (\ref{Tresgen})
$\displaystyle  \int_0^\infty \D t 
\langle \tau | t \rangle h(t) Q_0(X_r,t)$. The integrand is 
the mean of $\tau$ conditioned on the first reset time $t$ multiplied by the joint probability of first reset at $t$ and survival of the process up to time $t$. Thus the integral is simply the mean of $\tau$ conditioned on survival of the process up to $\tau$ multiplied by  the probability of survival of the process.

We may write the second term in (\ref{Tresgen})
as $\displaystyle \langle \tau | survival \rangle \frac{P}{Q}$
where $P$ is the probability of the process surviving up to the first reset
and $Q=1-P$.
Then, considering each reset period as a Bernoulli  trial which is successful if the process survives, the ratio of success to failure is the mean number of successful trials before a failure.
Thus 
\begin{eqnarray}
T(X_r)   &=&  T_{\tau=0}(X_r) + 
\displaystyle \frac{\int_0^\infty\D \tau  \langle \tau | t \rangle h(t) Q_0(X_r,t)}
{1- \int_0^\infty \D t\, h(t)  Q_0(X_r,t)}
\label{Tresgen2}
 \end{eqnarray}
where the second term on the rhs, which results from the occurrence of a refractory period is simply
the product of the mean duration of a refractory period after
a  reset period during  which the process survives with the mean number of successful reset periods before absorption.

\subsection{Poissonian reset}

Here let us return to the case of Poissonian reset with rate $r$ but now with a correlated refractory period
\begin{equation}
H(t, \tau) = r {\rm e}^{-rt} W(\tau|t)
\end{equation}
We  assume that one can express the conditional mean appearing in (\ref{Tresgen},\ref{Tresgen2})
 as a power series in $t$
\begin{equation}
\langle \tau | t \rangle = \sum_{n=0} a_n t^n\;.
\end{equation}
Then we can express the quantities in (\ref{Tresgen}) explicitly in terms of the Laplace transform of the survival probability without resetting and its derivatives
\begin{equation}
T(X_r)   =  \frac{\tilde Q_0(X_r,r)
+ r \sum_{n=0}(-1)^n a_n \tilde Q^{(n)}_0(X_r,r)}
{1- r \tilde Q_0(X_r,r)}\;.
\label{TresgenP}
 \end{equation}
For example, the case of independent refractory period
has $a_0 = \langle \tau \rangle$ and $a_n=0$ for $n>0$.
The example of Section \ref{sec:ex} has $\langle \tau \rangle = a_1 t$
with the the other coefficients $a_n =0 $ for $ n \neq 1$.




\section{Joint active time and first passage time distribution}
Finally we consider the active time of a particle
which is the time elapsed {\em excluding}  refractory periods.
The knowledge of the distribution of active  time and time spent in the quiescent state may be useful when one has to optimise activity. 
We can write down a  recursion relation for the joint probability density $P_r(t_a, t)$ of
the particle  being absorbed at time $t$ and having spent  active time $t_a$
under resetting dynamics (here we just consider  Poissonian resetting with
rate $r$ and independent refractory period).
The recursion relies on dividing up trajectories according to how many resets have occurred:
\begin{eqnarray}
P_r(t_a, t) &=& {\rm e}^{-rt} F_0(t) \delta(t-t_a) \nonumber \\
&+& r \int \D t_1 \D t_2 \D t_3   {\rm e}^{-rt_1} Q_0(t_1) W(t_2) {\rm e}^{-rt_3} F_0 (t_3)
 \nonumber \\
&& \times
 \delta(t-(t_1+t_2+t_3)) \delta(t_a-(t_1+t_3))\nonumber \\
&+& r^2 \int \D t_1 \D t_2 \D t_3 \D t_4 \D t_5   {\rm e}^{-rt_1} Q_0(t_1) W(t_2) {\rm e}^{-rt_3} W(t_4) {\rm e}^{-rt_5} F_0 (t_5) \nonumber \\
&& \times \delta(t-(t_1+t_2+t_3+t_4+t_5)) \delta(t_a-(t_1+t_3+t_5))\nonumber \\
& +& \ldots
\end{eqnarray}
The first term on the rhs is the contribution from  trajectories in which there is no resetting which occurs with probability ${\rm e}^{-rt}$. For these trajectories $t_a =t$ and given there is no resetting the survival probability is that of the process without resetting which we denote $Q_0(t)$. The second term 
is the contribution from trajectories with one reset which occurs at time $t_1$,
a refractory period of duration $t_2$ and a period  with no resetting of duration $t_3$ at the end of which the particle is absorbed which occurs
with first passage probability
of the process without resetting which we denote $F_0(t)$.
Similarly, the third term  is  the contribution from trajectories with two resets which occur at time $t_1$
followed by refractory period of duration $t_2$ 
and at $t_1 + t_2 + t_3$ followed by a refractory period of duration $t_4$, and  with no resetting of duration $t_5$ at the end of which the particle is absorbed. Clearly there is an infinite sequence of contributions corresponding
to higher numbers of resets.

Now taking the Laplace transform with Laplace variable $s$ conjugate to $t$ and Laplace variable $\lambda$ conjugate to $t_a$
\begin{eqnarray}
\tilde P_r(\lambda,s)
&=& \tilde F_0(\lambda+r+s)
+r \tilde Q_0(\lambda+r+s) \tilde W(s) \tilde F_0(\lambda+r+s)\nonumber \\
&&+r^2 \tilde Q_0(\lambda+r+s)^2 \tilde W(s)^2 \tilde F_0(\lambda+r+s)
+ \ldots
\end{eqnarray}

Summing the geometric series  yields
\begin{equation}
\tilde P_r(\lambda, s) = \frac{\tilde F_0(\lambda+r+s)}{1-r \tilde Q_0(\lambda+r+s) \tilde W(s)}\;.
\end{equation}

First let us consider the marginal distributions for $t$ (obtained by setting $\lambda=0$) and for $t_a$ (obtained by setting $s=0$)
\begin{eqnarray}
\tilde P_r(0, s) &=& \frac{\tilde F_0(r+s)}{1-r \tilde Q_0(r+s) \tilde W(s)}
\label{Plambda0}\\
\tilde P_r(\lambda, 0) &=& \frac{\tilde F_0(\lambda+r)}{1-r \tilde Q_0(\lambda+r)}\;.
\label{Ps0}
\end{eqnarray}
The first equation (\ref{Plambda0}) recovers (\ref{Qrs})
when we identify
\begin{equation}
\tilde P_r(0, s) = \tilde F_r(s) = 1-s \tilde Q_r(s)
\end{equation}
The second equation (\ref{Ps0}) shows that distribution of active time is precisely the first passage time distribution in the absence of resetting, as expected.

\section{Conclusion}
In this note we have considered the effects of  a cost for resetting in the form of a refractory period generalising some previous results \cite{RUK14,RRU15,R16}. We have shown  how calculations of the probability distribution and survival probability in the presence of an absorbing boundary may be carried out by using a renewal equation based on the first reset.
The calculations yield some interesting novel results:
(i) As a result of a non-zero refractory period the stationary state  develops a delta function at the 
resetting position with weight proportional to the mean refractory period
and how this peak emerges in the long-time behaviour. For a power-law distribution of refractory periods one can obtain slow (\ref{Bt.gamma>2},\ref{Bt.gamma<2}) power-law relaxation or ultra-slow
(\ref{Bt.gamma2}) inverse-logarithmic relaxation to the stationary behaviour.
(ii)  The survival probability and MFPT 
is strongly affected by correlated resetting and refractory periods.
(iii) A joint active time and first passage time distribution may be computed.

It should be straightforward to extend
our renewal approach to processes in higher dimensions in the presence of a
random refractory period.
The joint active time and mean first passage time distribution should allow one to calculate other quantities of interest, for example the mean first passage time conditioned on the active time.
 It would also be of interest to
see how the presence of a finite refractory period affects other
observables associated to reset dynamics~\cite{MSS15,R16,MMV17}.
Finally, the refractory period may have interesting consequences for
bacterial population dynamics in the presence of `catastrophic' events
that are similar to reset events~\cite{VAME}.

\ack
MRE acknowledges a CNRS Visiting Professorship and thanks
LPTMS for hospitality. SNM thanks the Higgs Centre for Theoretical Physics for hospitality.

\end{document}